# Vibrational Coupling to Epsilon-Near-Zero Waveguide Modes.

*Thomas G. Folland[1], Guanyu Lu[1], A. Bruncz[1,2], J. Ryan Nolen[3], Marko Tadjer[4] and Joshua D. Caldwell[1]*

1. Department of Mechanical Engineering, Vanderbilt University, Nashville, TN, USA, 37212

2. Department of Physics, University of Alabama in Huntsville, Huntsville, AL, USA, 35899

3. Interdisciplinary Materials Science, Vanderbilt University, Nashville, TN, USA, 37212

4. US Naval Research Laboratory, Washington, DC 20375, United States of America

*josh.caldwell@vanderbilt.edu

ABSTRACT: Epsilon near zero modes offer extreme field enhancement that can be utilized for developing enhanced sensing schemes. However, demonstrations of enhanced spectroscopies have largely exploited surface polaritons, mostly due to the challenges of coupling a vibrational transition to volume-confined epsilon near zero modes. Here we fabricate high aspect ratio gratings (up to 24.8 μm height with greater than 5 μm pitch) of 4H-SiC, with resonant modes that couple to transverse magnetic and *transverse electric* incident fields. These correspond to metal-insulator-metal waveguide modes propagating downwards into the substrate. The cavity formed by the finite length of the waveguide allows for strong absorption of incident infrared light (>80%) with Q factors in excess of 90, including an epsilon near zero waveguide mode with $\varepsilon_{eff}=0.0574+0.008i$. The localization of the electromagnetic fields within the gap between the grating teeth suggests an opportunity to realize a new platform for studying vibrational coupling in liquid environments, with potential opportunities for enhanced spectroscopies. We show that these modes are supported in anhydrous and aqueous environments, and that high aspect ratio gratings coherently couple to the vibrational transition in the surrounding liquid.

The electromagnetic field confinement offered by surface polaritons[1-2] and epsilon-near-zero (ENZ)[3-7] modes have long been discussed for applications in surface-enhanced sensing[8-13], and vibrational coupling[14-15]. Whilst surface plasmon polaritons (SPPs) have been extensively explored, demonstrating enhanced spectroscopies using ENZ modes has remained challenging. This is largely because at optical frequencies ENZ modes are often realized by coupling light into a material where epsilon is close to zero[16-18], or via waveguides that are not hollow and therefore incompatible with confining the analyte of interest within the region of highly confined electromagnetic fields[3-4]. In this letter we investigate high-aspect-ratio grating (HAG) structures designed to support surface phonon polaritons (SPhPs)[1] at the interface between the polar crystal grating surfaces and the surrounding environment[19]. We show that the modes supported by this structure behave like metal-insulator-metal (MIM) waveguide modes in a short cavity[20-22]. Furthermore, due to this architecture, these structures support an ENZ mode in the gap between the grating teeth. This enables the first colocation of strongly confined ENZ fields with an analyte of interest, including liquids, with large surface area. As proof of this, we demonstrate that the ENZ fields can coherently couple to vibrational transitions in a liquid. Thus, this constitutes a platform for studying ENZ and SPhP strong coupling at infrared (IR) frequencies, with potential applications in surface enhanced spectroscopies[8-13, 23] as well as light-controlled chemistry[14-15].

This study exploits high aspect ratio gratings, which have a height ($h$) that is much larger than their period ($\Lambda$) (see Fig 1). In HAGs and nanopillars, surface polariton modes are supported between the teeth, propagating as MIM waveguide modes downwards into the grating[20-22, 24-25]. The frequency of the modes can be controlled by changing the effective index of refraction ($n_{eff}$) of the polariton wave using the size of the air gap ($g$) between the teeth, or the height of the grating ($h$, see Fig. 1 a-c). Furthermore, polaritonic modes in these structures have been demonstrated to improve sensitivity for surface-enhanced Raman (SERS) and solid-phase IR sensing[26-27]. Existing studies have focused on transverse magnetic (TM) waveguide modes in materials supporting surface plasmons, as opposed to the transverse electric (TE) waveguide modes that can support ENZ type behavior. In our work, we leverage SPhPs in 4H-SiC, which can be supported within the spectral region between ~792 and 972 cm$^{-1}$. This spectral region is referred to as the Reststrahlen band and is bound by the transverse (TO) and longitudinal optic (LO) phonons of the material[28]. The low scattering losses associated with polar optic phonons[19, 29], grating[30-32] and nanoantenna[28, 33-36] structures can support resonances with high Q factors, with a maximum reported of ~300[28, 33] and 400 for far-field and near-field[37] measurements, respectively. This makes them an ideal system for studies of ENZ materials and enhanced spectroscopic methods.

Despite the impressive properties of SPhPs, there have been limited demonstrations of these modes for realizing effects such as surface enhanced infrared absorption (SEIRA)[8-13, 23] spectroscopy, especially on non-solid analytes[38]. This is largely due to the challenges of coupling to the highly localized fields supported by SPhP resonators in a liquid environment, where infrared light is generally strongly attenuated. The waveguide modes in HAGs discussed here overcome this problem, as the optical fields are strongly localized within the gaps between grating teeth. One of the remarkable properties of the HAG structures studied in this work is that their resonant modes couple to both TM and TE incident fields[39]. Whilst TM incident fields form confined MIM polaritonic modes (with $n_{eff}>1$, as in past work[30-32]), the TE-coupled modes exhibit behavior similar to near-cutoff frequency of waveguides[3-6], showing $n_{eff}<1$. Here, we provide a complete analytical model for these TE modes, and demonstrate that they can be attributed to standing waves in a finite length MIM grating. This allows us to demonstrate that the lowest-order TE mode can be classed as an ENZ waveguide mode, with an $\varepsilon_{eff}=0.0574+0.008i$. In order to explore the possibility of leveraging these ENZ modes for liquid-based sensing, we study the properties of the resonant modes in various working liquids including water and alcohols. We find that the MIM modes are supported in anhydrous environments and can also be observed in aqueous environments, but are attenuated and broadened in the latter. We then use mixtures of different solvents to study the possibility of surface-enhanced sensing and strong coupling within these structures. We show that waveguide modes in HAG gratings can couple to vibrational transitions in the liquid, resulting in mode splitting indicative of strong coupling, with a splitting of up to 7.8 cm$^{-1}$ about the ENZ resonant frequency. Our results illustrate that HAG structures provide a platform for studying vibrational coupling between SPhPs and vibrational bands in the IR[14-15, 40-42], and constitutes the first implementation of ENZ modes for coupling to vibrational bands in liquids.

Gratings with a pitch ($\Lambda$) ranging from 5-10 μm and grating tooth spacings ($g$) (see Fig. 1) such that a constant fill fraction (g/$\Lambda$) of *0.5* is maintained and were fabricated using a combination of photolithography and deep-reactive ion etching (see methods). Three different etch depths (Fig. 1a-c) were used, and the samples were characterized using FTIR micro-spectroscopy (see methods). The shallow grating structures (0.8 μm) exhibit a series of SPhP grating modes when reflection is collected in TM polarization (incident field oriented along $E_x$), as shown in Fig. 1(d), and explored in past works[30-31]. However, no discernable modes were observed in TE polarization (field was oriented along $E_y$), as anticipated for a grating featuring a sub-diffractional pitch (Fig. 1e). For deep structures (11.5 and 24.8 μm), TM polarization also resulted in the excitation of SPhP modes, albeit with additional absorption bands. Interestingly, the TE polarized reflection spectra of these deep-etched gratings also exhibit strong resonant response, with a series of sharp absorption peaks observed throughout the SiC Reststrahlen band (Fig. 1). In contrast, as shown in the shallow gratings, TE polarized light cannot directly excite SPhP modes, while the narrow grating pitch prevents in-coupling via diffractive orders with normal incidence light. Thus, the modes in our HAG structures must be derived from waveguide-like modes, which are known to be supported in MIM structures[4,6]. While the SiC gratings discussed here do not contain any metal, the negative permittivity of the SiC within the Reststrahlen band provides the metallic behavior necessary for the MIM design. Numerical simulations of the grating spectra support this ascertion, as shown in Supplementary Section 1, and Supplementary Fig. 1. Both TE and TM field profiles illustrate electromagnetic fields confined within the gap between the grating 'teeth', with a series of standing waves in the cavity. The localization of the electromagnetic fields within the teeth is ideal for enhanced spectroscopic applications, as the air gaps can be filled with a material for analysis. Furthermore, the TE waves have a longer wavelength than light in free space, which suggests an extremely low refractive index, and hence an indication of ENZ behavior.

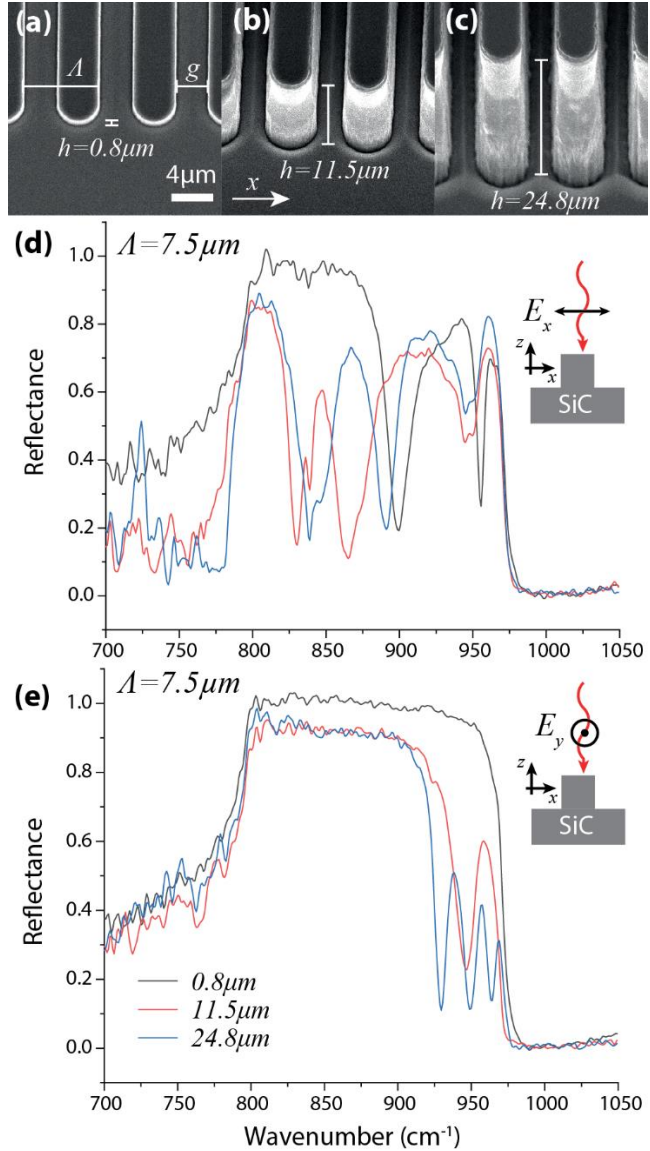

Fig 1. Scanning electron microscope (SEM) images of deep-etched SiC grating structures (a)-(c) illustrate the significant variation in etch depths ($h$ = 0.85, 11 and 24.8 μm, respectively) of three grating structures with $\Lambda$ = 7.5μm and $t$ = 4μm as viewed in an electron microscope at 30˚ off normal. Polarized FTIR spectra with electric field in (d) TM or (e) TE orientation.

To provide a complete description of the TE-polarized waveguide modes and their ENZ properties, we can study the pitch dependence of the modes (Fig. 2a) and compare to a simple analytical model for a cavity formed in a short waveguide (Fig. 2b/c). The narrowest pitch structure supports modes located at the high-frequency edge of the Reststrahlen band. As the grating pitch is increased, the modes continuously red-shift, and additional modes appear. This is consistent with a continuous tuning in the dispersion of the waveguide modes as the gaps between the SiC teeth increases. To demonstrate this we use a simple slab waveguide model to calculate the dispersion of the TE-polarized waveguide modes (inset Fig. 2b), employing the standard MIM waveguide equations for symmetric and antisymmetric modes[6];

$$k_2 + k_1 tanh\left(\frac{-ik_1 g}{2}\right) = 0 \quad (1)$$

$$k_2 + k_1 coth\left(\frac{-ik_1 g}{2}\right) = 0 \quad (2)$$

Here, $g$ is the thickness of the gap, $k_i^2 = \varepsilon_i k_0^2 - \beta^2$, $\varepsilon_i$ is the dielectric function of each layer (1= air, 2 = SiC), $k_0$ is the free-space wavevector and $\beta$ is the guided wavevector. Instead of plotting the waveguide propagation constant $\beta$, we choose to plot the effective index of the wave, ($n_{eff} = \beta/k_0$), as shown in Fig. 2b. We find that the cutoff frequency of the TE waveguide mode (where $n_{eff}$ goes to zero) varies dramatically as the pitch is increased, exhibiting a significant red-shift, consistent with our experimental data. Furthermore, to more accurately compare to our experimentally measured modal frequencies, we can also express a cavity resonance condition in terms of $n_{eff}$[24]. This condition will state at which values of $n_{eff}$ the cavity wave is supported in the continuous dispersion spectrum, for a given cavity length. The fields of Fig. S1 suggest that the waveguide forms a closed cavity, from which we can write the closed cavity condition:

$$n_{eff} = \frac{l\lambda}{2h} \quad (3)$$

where $l$ is the mode order, $\lambda$ is the free-space wavelength and $h$ is the grating height. The cavity condition is plotted in Fig. 2b using dashed lines. The points where the cavity condition (dashed) and waveguide dispersion (solid) lines intersect indicate where we would anticipate experimentally measured modes (symbols). In general, the experimental modal positions qualitatively agree in the trends and rough spectral shape. Deviations in the mode frequencies are likely due to two effects – lateral coupling between different waveguide teeth (evident by examining fields presented in Fig. S1), and tapering of the waveguide width that occurred due to the RIE processing, neither of which is considered in the analytical model. However, this simple model does provide validation that these resonances can be attributed to TE-type waveguide modes with $n_{eff}$<1, and demonstrates that a simple model can be used to estimate the mode frequencies for a simplified design process. We note that the *l=1* mode possesses an extremely low effective mode index of 0.235, which equates to an effective permittivity of $\varepsilon_{eff}$=0.0574+0.008i. As this is a comparable permittivity to exemplary ENZ materials, including doped CdO[16] ($\varepsilon_{eff}$~0+0.4i) and AlN[17] ($\varepsilon_{eff}$~0+0.05i), the waveguide modes at the heart of this work can also be considered an ENZ waveguide mode.

This same cavity model can also be used to estimate the modal *Q* factors associated with these frequencies. We can write the *Q* factor as;

$$Q = \frac{\tau\omega}{2} \quad (4)$$

where $\tau$ is the modal lifetime, and $\omega$ is the frequency of the mode. The modal lifetime can be written as a combination of the lifetime associated with cavity ($\tau_c$) and associated with loss from the MIM mode ($\tau_s$)[43]:

$$\frac{1}{\tau} = \frac{1}{\tau_c} + \frac{1}{\tau_s} = v_g \left( \frac{\ln(R)}{2h} + \text{Im}(\beta) \right) \quad (5)$$

where $v_g$ is the group velocity of the mode (calculated from the dispersion) and *R* is the power reflection coefficient at the top of the cavity $R = \left( (n_{eff} - 1)/(n_{eff} + 1) \right)^2$. The calculated *Q* factor shows a sharp increase close to the cutoff frequency of the mode as shown in Fig. 2(c) - attributed to the high cavity reflection as $n_{eff}$ and $\varepsilon_{eff}$ approach zero. The cavity reflection drops off rapidly away from this frequency, and this limits our modal lifetime at higher frequencies (see Supplementary Fig. S2). The experimentally measured $Q$ values are similar to those predicted by the model, but we find the analytical model overestimates $Q$ for the $l=1$ mode, and underestimates it for $l=2$ and $3$. This is again likely due to the non-vertical sidewalls in our structures, which will modify the dispersion of Fig. 2(b) and hence change the MIM mode group velocity, on which the calculated $Q$ is directly dependent. However, the similarities between experimentally measured and analytically predicted $Q$s provides further evidence that these modes can be seen as standing waves in a finite length MIM waveguide. These trends are also observable in intermediate aspect ratio ($h=11.5\mu m$) gratings, as shown in Supplementary Fig. S3.

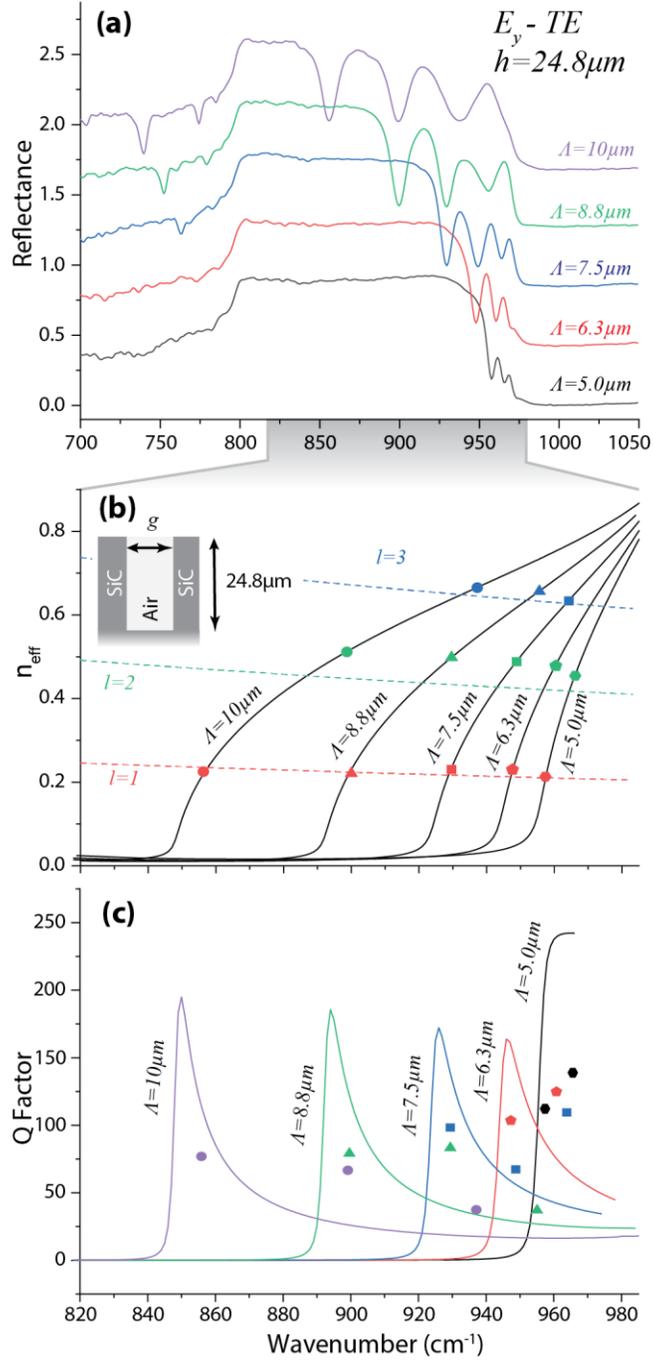

Figure 2. Pitch tuning of TE type waveguide modes in deep SiC gratings, (a) experimentally measured by FTIR spectroscopy. The absorption below the TO phonon frequency is due to high index modes forming in the SiC teeth. (b) Analytical mode solutions for the waveguide (model inset). Mode positions of standing waves formed by the TE waveguide modes can be found from the crossing of the waveguide dispersion (solid lines), and the cavity resonance condition (dashed lines, colored). The experimentally measured mode positions are overlaid on top of the analytical

dispersion curves in (b), with circles, triangles, squares, pentagons and hexagons representing modes in 10,8.8,7.5,6.3 and 5.0μm gratings respectively. (c) shows the analytically calculated Q factor compared against the experimentally measured Q factors, with the same shapes as (b).

Given the ability of such structures to create resonant cavity modes (including an ENZ mode), we can now turn our attention to their application for liquid sensing. In particular, the Reststrahlen band of SiC lines up with the frequencies of many volatile organic compounds, where high sensitivity sensing is of distinct interest[23]. To study the potential for using HAG gratings in liquid sensing applications, we first study their mode spectra in liquid environments. To achieve this, we placed the grating samples inside a liquid cell (see methods) compatible with the FTIR spectrometer (Fig. 3a) to probe the response in the presence of a liquid (water and acetone). We measured the IR reflectance spectrum from the grating structures (Fig. 3b), as well as from the un-patterned surrounding substrate to the side of the gratings, which provides a 'reference' measurement of the strength and frequency of the unperturbed absorption bands from the liquid itself. For acetone, the vibrational band occurs at ~900 cm$^{-1}$, whereas for water light is generally attenuated across the complete Reststrahlen band of SiC. When submerged in acetone, the grating exhibits two additional waveguide modes, which can be attributed to the higher index of the surrounding medium modifying the waveguide dispersion. However, the modes have similar Q factor and absorption to the grating in air. In water the grating spectrum is heavily attenuated, but has enhanced transmission versus the spectrum taken through liquid alone. This is likely due to the localization of the electromagnetic fields inside the silicon carbide, which reduces loss from the liquid. These results demonstrate that HAG structures are compatible with liquid-based sensing approaches, and in supplementary section 4 we show that the frequency tuning is comparable to past demonstrations of liquid sensing[8].

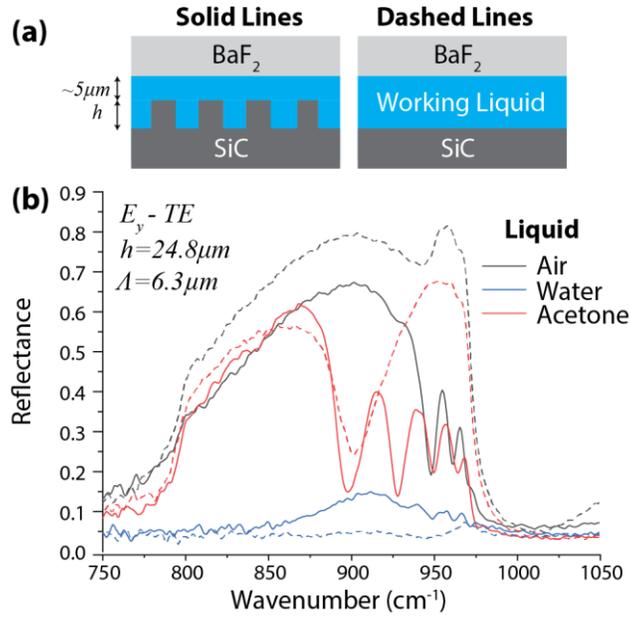

Figure 3. Response of TE waveguide modes in a liquid environment (a) shows a schematic of the experimental configuration, and (b) mode spectrum of the HAG inside a liquid cell filled with air, water and acetone

To investigate the possibility of using HAGs and ENZ modes as a platform for vibrational sensing and strong coupling, we also studied the behavior of polaritonic resonances in mixtures of a range of different solvents, including acetone, IPA, toluene and cyclohexane, all of which have some degree of vibrational activity within the SiC Reststrahlen band (See Fig. S5). As cyclohexane exhibits a relatively strong (absorption coefficient estimated at 334 cm$^{-1}$), sharp absorption band ($CH_2$ rocking) located at the center of the Reststrahlen region (903 cm$^{-1}$), which makes it ideal for such studies. In these studies, we focus on the TE-type modes, as the mode spectrum for these structures is significantly simpler.

In order to tune the cavity mode to the frequency of the cyclohexane vibration, we study a range of different gratings with slightly different pitches, as shown in Fig. 4(a). We find that a grating with a $\Lambda$=6.48 μm results in an ENZ waveguide mode with strong spectral overlap with the vibrational mode of the neat cyclohexane that it is immersed in. The strong coupling between these two resonant modes results in a modal splitting, which is indicative of strong coupling[14, 41-42]. In order to verify that this can be attributed to coherent coupling, we dilute the cyclohexane and study how the modal splitting changes with modified molecular concentration. However, to ensure that the cavity mode of the ENZ waveguide is not spectrally shifted away from the vibrational resonance, a solvent with a similar refractive index as cyclohexane must be used. We find that a 1:1 mixture of IPA and toluene serves as an acceptable solvent as it maintains the cavity resonance frequency close to the 903 cm$^{-1}$ vibrational band of cyclohexane (see Fig. S6). Varying the concentration of cyclohexane in this solvent mixture demonstrates a tunable splitting between the waveguide mode and the vibrational transition (Fig. 4b). By fitting the data in Fig. 4(b) we show that the mode splitting follows the simple $\sqrt{N}$ dependence on the concentration of molecules (Fig. 4c)[14], providing further evidence of vibrational strong coupling. The maximum mode splitting is 7.8 cm$^{-1}$, which is comparable to the FWHM of the uncoupled cyclohexane vibrational resonance (7.9 cm$^{-1}$) and the cavity mode (14.5 cm$^{-1}$), placing the observed splitting on the edge of the threshold of strong coupling. By implementing a crude dielectric function that would be analogous to cyclohexane, we see that this result is also consistent with our analytical model (See Fig. S7). We note that whilst IPA could in principle be used as an excellent target molecule as it possesses a very strong absorption band at 952 cm$^{-1}$, with $\alpha$=3300 cm$^{-1}$. However, in practice, the vibrational energy of the mode is too close to the edge of the Reststrahlen band, as shown in Supplementary Figs. S8, S9 and S10, and as a result the mode is only broadened. This highlights the importance of using the *l=1* ENZ mode (which exhibits the largest Q) for such strong coupling experiments.

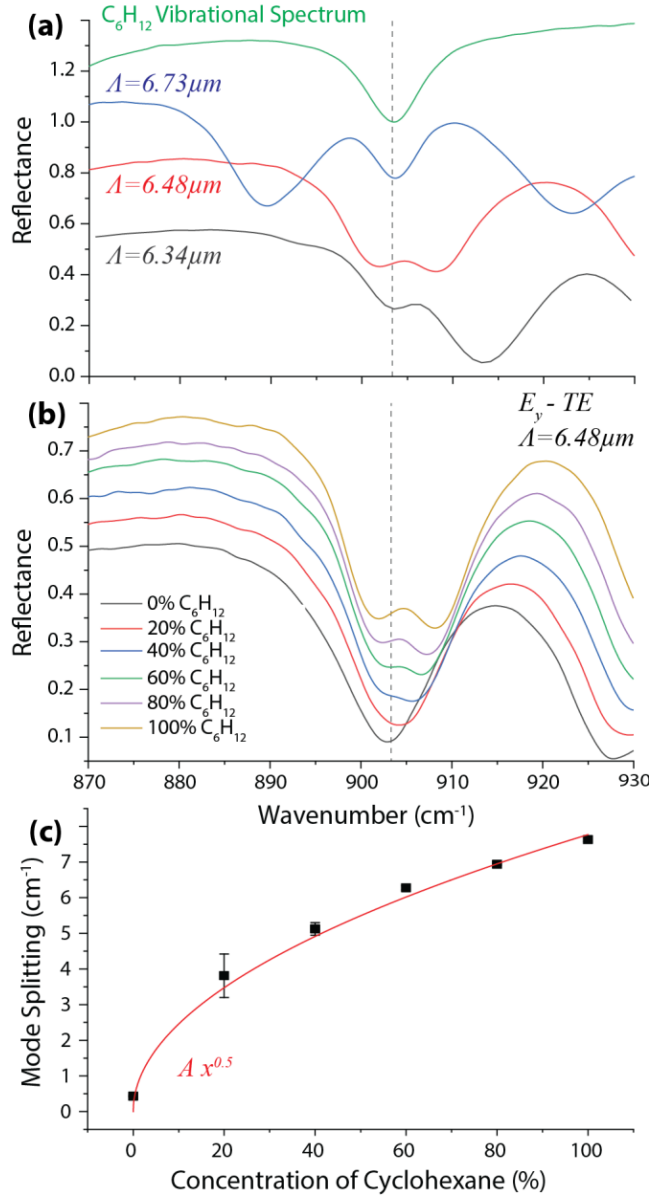

Fig. 4 Strong coupling between a waveguide mode and the vibrational mode of Cyclohexane. (a) shows how the grating pitch can be controlled in order to overlap the cavity mode with the vibrational mode, producing to mode splitting. (b) concentration dependent mode splitting in HAG modes. (c) concentration dependent mode splitting in the vibrational cavity.

In summary, we have fabricated HAGs made of 4H-SiC, which supports both TE- and TM-polarized SPhP waveguide modes in the long-wave IR. Unlike the response from a shallow grating, where modes propagate along the surface of the grating teeth, in HAG structures the modes propagate vertically downwards into the substrate. These waveguide modes are generally described by simple MIM waveguide equations, and strongly localize light in between the grating teeth. For TM modes, this is associated with polaritonic confinement effects, however for TE modes this is associated with a reduction in the effective mode index to values less than 1. Due to the finite height of the grating, the modes formed in the gratings take the form of standing-wave-type cavity modes, with the lowest order resonance behaving as an ENZ waveguide mode. The frequency of the modes is generally described by a combination of the waveguide dispersion, and a cavity resonance condition. Finally, we show that these modes can be supported in liquid environments – finding that whilst SPhP modes are readily supported in anhydrous environments, they are heavily attenuated in aqueous environments. Finally, we show that these gratings can support vibrational strong coupling between the ENZ mode and the vibrational band in cyclohexane. This constitutes the first implementation of ENZ modes for enhanced spectroscopy, as well as the formation of a hybrid phonon-vibration polariton mode.

**Methods**

**Device Fabrication**

Gratings with a pitch ($\Lambda$) ranging from 5-10 μm and grating tooth spacing ($g$) (see Fig. 1) with a constant fill fraction ($g/\Lambda$) of $0.5$ were fabricated using a combination of photolithography and inductively-coupled plasma (ICP) process. Specifically, an on-axis, undoped 4H-SiC wafer was cleaned in piranha solution, followed by a DI rinse and $N_2$ blow dry. The wafer was seeded by Cr/Au (100 Å/1000 Å) deposited using electron-beam evaporation, patterned using contact photolithography, and selectively electroplated with a Ni etch mask (1.7 μm thick). Dry etching was carried out following the processes developed in Ref. [44]. Following etching, the metal mask was removed in aqua regia for 1 hour, followed by 60 seconds in chrome etchant and a DI rinse.

We show scanning electron images of the fabricated grating structures with $\Lambda$ = 7.5µm and $g$ = 3.5µm in Fig. 1(a)-(c) with three different etch depths; 0.8 µm, 11 µm and 24 µm. These depths were obtained using a 7 minute β3-etch (1:9 $SF_6:O_2$, 50 sccm total flow), a 60 minute modified β-etch with $SF_6:O_2$ ratio reduced to 1:6 for better mask selectivity, and a 60 minute standard α etch with 10:1 $SF_6:O_2$ ratio respectively. Measurements of the profile of all gratings studied show an average sidewall angle of approximately 2.3°±1.2°.

**Optical Characterization**

Reflectance spectra for the grating structures were measured using a Bruker Hyperion 2000 IR Microscope coupled to a Bruker Vertex 70v Fourier transform IR spectrometer (FTIR) using a spectral resolution of 2 $cm^{-1}$. The microscope is equipped with a refractive Ge objective (5x, NA=0.17, Pike Technologies), a Ge wire grid polarizer (Pike Technologies) and liquid-nitrogen-cooled mercury cadmium telluride (MCT) detector. The use of a 5x objective removes out-of-plane electromagnetic field components from the incident field, which simplifies the analysis of our initial results. It also ensures that all of the structures studied are well-below the diffraction limit (removing conventional diffractive effects).

**Numerical Simulations**

Simulations were performed in CST studio suite 2018, where the SiC grating structure uses a single grating period and unit cell boundary conditions, and a perfectly matched layer for the substrate. The dielectric function used for 4H-SiC was derived from that presented in Ref. [45]. In all of our analyses we define the periodic axis of the grating as the x-axis, and the etch direction as the z-axis (see Fig 1.)

**Liquid Based Experiments**

Liquid experiments were performed in a Pike Technologies demountable liquid cell using $BaF_2$ windows. To use the liquid cell with our small (1cm by 1cm by 500μm) samples, a 0.02' (508μm) Teflon spacer was used in the cell. Two samples could be loaded in the cell simultaneously, allowing accurate comparison between experimental results on different devices. The cell was loaded through a drilled $BaF_2$ window using a syringe, and the liquid was changed by drying the liquid cell using dry nitrogen. The Teflon spacer was not the exact thickness of the SiC wafer, but by examining the fringes formed by the $BaF_2$-air-SiC cavity outside the Reststrahlen band when an FTIR spectrum is taken next to the grating (See Supplementary Fig. S11), we can estimate the spacing between the cover and the sample. The frequency spacing of the fringes ($\Delta f$) 673 cm$^{-1}$ (197 cm$^{-1}$), for *h=0.8μm* (*h=24.8 μm*) gratings is related to the size (*L*) of the cavity by *L=c/(2nΔf)*, where c is the speed of light and n is the index of the medium. This gives a height of the control cavity of 7.4μm for the grating with *h=0.8 μm*, and 25.4μm for the grating with *h=24.8 μm* – suggesting that the cell provides a gap of a few microns between the top of the grating and the $BaF_2$ cover.

AUTHOR INFORMATION

**Corresponding Author**

Prof. Joshua Caldwell - Department of Mechanical Engineering, Vanderbilt University, Nasvhille, TN, USA, 37212 - josh.caldwell@vanderbilt.edu

**Author Contributions**

The manuscript was written through contributions of all authors. TGF and JDC conceived of the experiment. Samples were prepared by MT, and characterized by TGF and JRN, and AB. TGF and GL performed numerical simulations and analytical calculations. All authors have given approval to the final version of the manuscript.

**Funding Sources**

Support for J.D.C., T.G.F and J.R.N was partially provided by the Office of Naval Research under contract number (N00014-18-1-2107 )/ J.R.N, T.G.F. and J.D.C. also acknowledge support from Vanderbilt School of Engineering through the latter's startup funding package. AB acknowledges support from the Vanderbilt University VUSE summer internship program. Work at the Naval Research Laboratory is supported by the Office of Naval Research.

ACKNOWLEDGMENT

We would like to thank Mr. Milton Rebbert for additional sample processing at NRL. We would also like to thank the Vanderbilt Institute of Nanoscale Science and Engineering for access to scanning electron microscopy

ABBREVIATIONS

SEM, Scanning Electron Microscopy; SPhP, Surface Phonon Polariton; ICP, inductively coupled plasma; HAG, high aspect-ratio grating; TE, Transverse electric; TM, transverse magnetic; TO, tranverse optical; LO, longitudinal optical; MIM, metal-insul; ator-metal; SEIRA, surface enhanced infrared absorption; IPA, isopropyl alcohol; ENZ, Epsilon Near Zero

# Supplementary materials: Vibrational Coupling to Epsilon-Near-Zero Waveguide Modes.

*Thomas G. Folland[1], Guanyu Lu[1], A. Bruncz[1,2], J. Ryan Nolen[3], Marko Tadjer[4] and Joshua D. Caldwell[1]*

1. Department of Mechanical Engineering, Vanderbilt University, Nashville, TN, USA, 37212

2. Department of Physics, University of Alabama in Huntsville, Huntsville, AL, USA, 35899

3. Interdisciplinary Materials Science, Vanderbilt University, Nashville, TN, USA, 37212

4. US Naval Research Laboratory, Washington, DC 20375, United States of America

*josh.caldwell@vanderbilt.edu

1. **Determining mode properties using numerical simulations**

To determine the properties of the resonant modes that were observed in the highest aspect ratio 4H-SiC gratings ($h$=24μm), we employed electromagnetic simulations, as shown in Fig. S1. For the narrowest period structure with $\Lambda$=5 μm (Fig. S1 a/b), numerical simulations reproduce the experimental results for both field orientations. The simulated electric field profiles for the four lowest frequency modes in the $E_x$ (TM) polarization (Fig. S1 i-iv), are all anti-symmetric across the gap. with a series of nodes oriented vertically downwards into the grating. Each progressive resonance exhibits an increase in the number of nodes (6, 7 and 8 for Fig. 2 i, iii, and iv, respectively). The feature at frequency 836 cm$^{-1}$, Fig. S1 ii, is not associated with a resonant mode (hence the similarities with Fig. 2iii), but instead arises from absorption from a zone-folded LO phonon in 4H-SiC[1], one of the unusual features of the vertical orientation of the field in these structures[2-3]. Finally, the absorption observed outside the Reststrahlen band is associated with waveguide modes concentrated inside the SiC, which has an extremely high index of refraction (up to $n$~20) at these frequencies due to the proximity to the strong absorption from the TO phonon[4]. In all, these simulations illustrate that the SiC grating forms a metal-insulator-metal TM polaritonic waveguide[5], and the vertical height of the grating forms a cavity resonator, consistent with earlier results in noble metals[6-8]. These waves show remarkably strong absorption at various frequencies across the entire Reststrahlen band, with a maximum absorption of 80% and $Q$~92 observed at 883 cm$^{-1}$, comparable with that achieved in many SPhP nanoantennas[3, 9-11]. However, unlike SPhP-based nanoantennas, in these HAG devices discussed here, the electromagnetic fields are not localized only at the corners of the structure, instead being distributed in the gap between adjacent grating teeth. This enables the modes to have a strong spatial overlap with local dipoles, such as molecular vibrational resonances, thereby providing further motivation to use these structures for liquid-based sensing modalities.

Whilst the plasmon polariton resonances occurring in HAG structures have been studied in prior work[6-8, 12-13], modes stimulated with $E_y$ (TE) polarization have not been discussed. The electromagnetic fields associated with the three modes for the $\Lambda$=5μm grating (Fig. S1 vi-vii) are a series of standing-wave resonances within the cavity formed by the grating. These field profiles are consistent with the formation of TE waveguide modes, with a wavelength $\lambda$~2×$h$~50μm for the lowest order wave, much longer than that of light in free space (approx.. 4.2x longer). This corresponds to an effective index ($n_{eff} = \lambda_0/\lambda$ where $\lambda_0$ is the free space wavelength), of less than one. Traditional dielectric and polaritonic waveguide modes typically have an $n_{eff}$>>1, but $n_{eff}$<<1 can occur under certain conditions in metal-insulator-metal waveguides. As the waveguide mode approaches the cutoff frequency, $n_{eff}$ of the waveguide mode drops to zero, also reducing the effective dielectric constant to zero. This leads to a significant enhancement of the electric field, as has been well studied in past work on epsilon near zero modes[5, 14-16]. Field enhancement can also be observed in our simulations - the peak electromagnetic field in the gap is stronger for the l=1 (by a factor of 20%) compared with the l=3 resonance. The fundamental TE mode at 957 cm$^{-1}$ exhibits $Q$~108 and ~65% absorption, giving it properties comparable to those induced with TM polarization (absorption of 80% and $Q$~92). Enhancement of the local electric fields in both polarizations demonstrates that HAGs offer different types of waveguide modes that can be exploited for field enhanced sensing, among other applications.

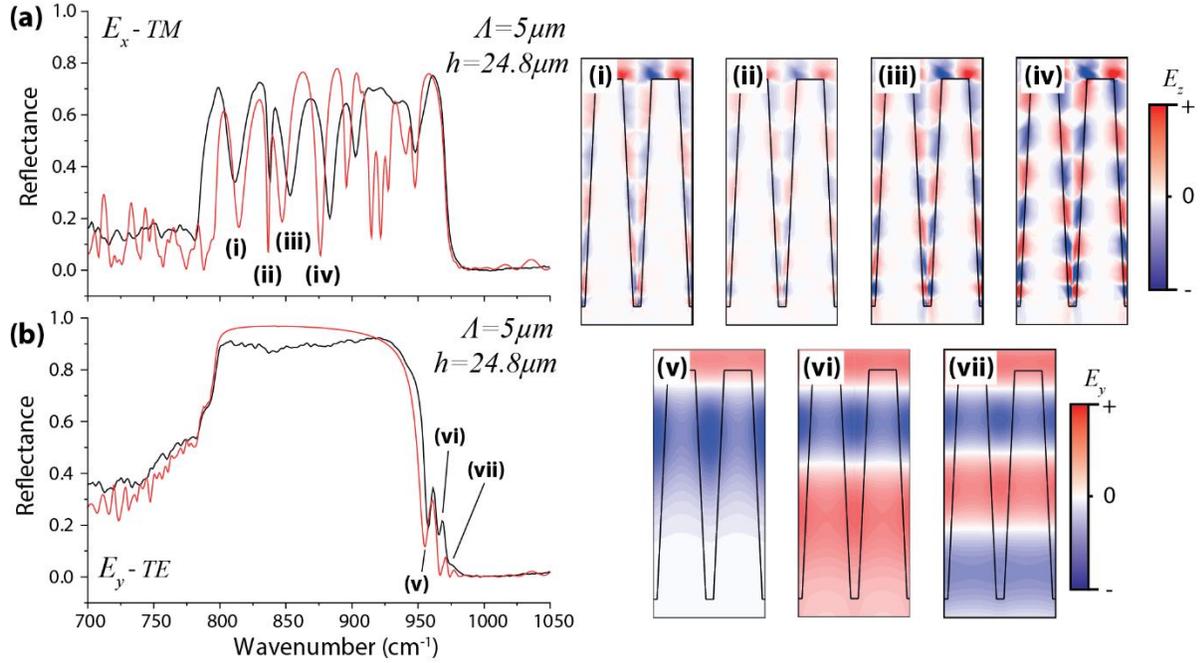

Figure S1. Waveguide modes in deep-etched SiC gratings. (a)/(b) shows TM/TE FTIR spectra for a grating of $\Lambda=5$ µm compared with numerical calculations. (i)-(vii) show the corresponding field profiles

2. **Mode Lifetime of TE waveguide modes**

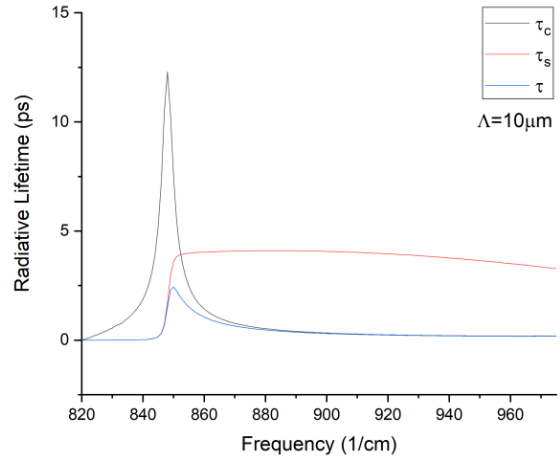

Figure S2. Mode lifetime ($\tau$) and contribution from the cavity $\tau_c$ and mode loss $\tau_s$ for the $\Lambda=10$µm, $h=24.8$µm grating

3. **Mode spectrum for $h=11.5$µm grating.**

In the main text we primarily discuss the mode spectrum of the deepest ($h=24.8$µm) grating. However, similar properties can be measured for the shallower, $h=11.5$µm gratings, and the same general physics applies. For example, shorter gratings still support TE waveguide modes, albeit with fewer resonances due to the limited length of the cavity (Fig. S1). These modes show similar tuning, albeit with lower quality factors and absorption depths. As such, height becomes a means to tune the frequency of TE waveguide modes, in addition to etch depth.

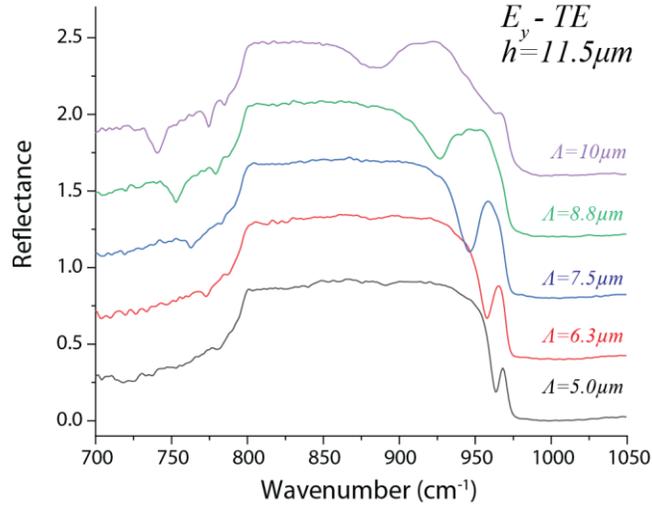

Figure S3. TE waveguide modes in a *h*=11.5μm grating. Fewer modes are observed than in Fig 2, due to the shallower depth of the grating.

4. **Comparison of spectra in liquid environments**

In addition to the studies of the TE modes in HAG structures, we also performed a comparative analysis with TM and grating waveguide modes. Through comparison of the IR reflectance spectra of the grating (solid lines) and unpatterned region (dashed lines) in different working liquids, the influence of the ENZ waveguide interactions with the liquid can be ascertained (Fig. S4 b-d). In the absence of a liquid, the grating resonances are clearly observable for both shallow and deep grating structures through the $BaF_2$ cover glass, similar to the results presented in Fig. 1-3. Upon submersion in acetone, the spectra for both the shallow ($h = 0.8 \mu m$, Fig. S4b), and deep gratings ($h = 24.8 \mu m$, Fig. S4c/d) develop multiple additional absorption bands that can be attributed to either polaritons, or absorption bands in the liquid. Whilst the polaritonic resonances are red-shifted due to the change in the local index when compared to their spectral location in air, the general features of the spectra are still present. The shifting of the modes suggests that these grating structures could also be attractive for index-based sensing in anhydrous environments. To assess the sensitivity to refractive index we can examine the shift in frequencies for (1) the TM mode in the shallow grating, and (2) the TE modes in the deep grating structures. Based on the 47 cm$^{-1}$ spectral shift of the lowest frequency TM mode in the shallow grating, and 50 cm$^{-1}$ shift of the lowest frequency TE mode in the deep grating, we estimate (assuming $n_{acetone}$ ~1.4[17]) a tuning sensitivity of 118 cm$^{-1}$/RIU and 125 cm$^{-1}$/RIU respectively. As the resolution of most commercial FTIR spectrometers is limited to a range 0.1-0.5 cm$^{-1}$ we can argue that the minimum detectable change in index is on the order of $8\times10^{-4}$ to $4\times10^{-3}$, comparable to past work using a prism to couple to SiC SPhPs[18]. This suggests that while there may be advantages to the HAG structures for index sensing, the advances are minor in comparison to index sensitivities in visible plasmonic systems. In water, despite the strong IR absorption, polaritonic resonances are still observed in both shallow and deep gratings. For shallow gratings the lines are significantly broadened in aqueous environments (from 15 cm$^{-1}$ to ~40 cm$^{-1}$ for the mode at ~884 cm$^{-1}$). However, for the HAG gratings we actually observe an enhanced optical signal from the grating structures vs the 'control' measurement. This is because much of the electromagnetic energy is actually stored within the low-loss SiC grating teeth, which increases the propagation length in water.

This alludes to the possibility of using SPhP structures to get longer optical path lengths in highly attenuating media. However, the resonances in such structures are also broadened, which makes it unclear if a HAG outperforms its shallow counterpart.

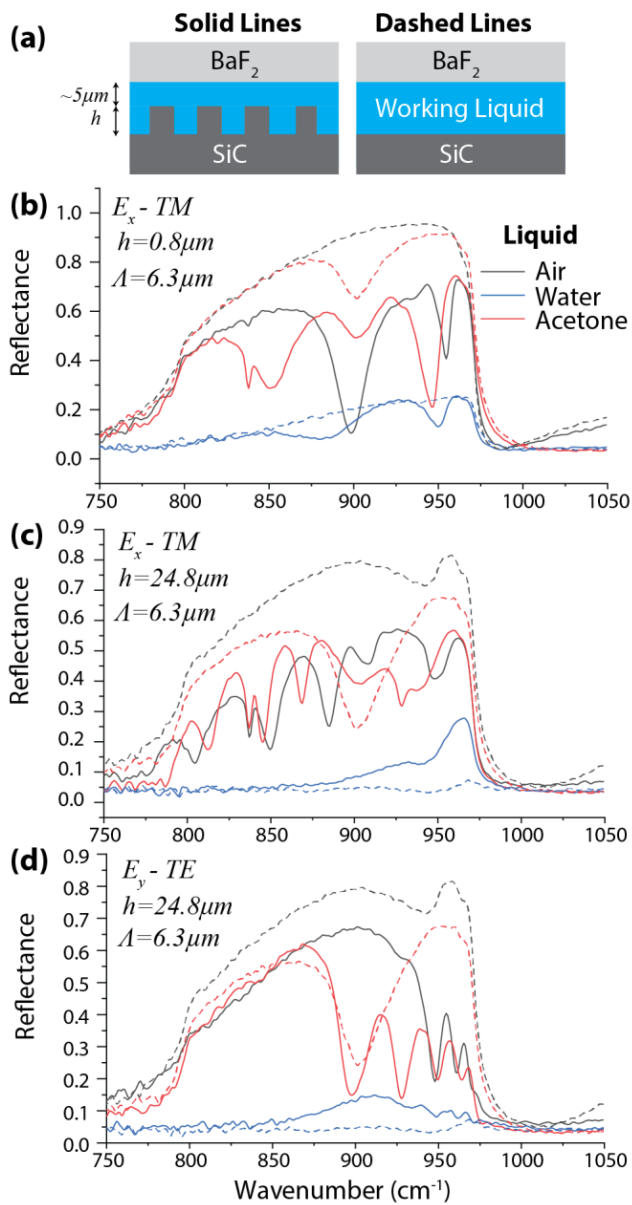

Fig. S4. Influence of liquid environment on grating resonances. A schematic of the experimental setup is shown in (a). FTIR reflectance spectra are shown in (b) and (c) for a shallow and deep grating respectively, with the electric field polarized along $E_x$, launching polariton modes. In (d) the we show reflectance spectra the deep grating with the electric field aligned along $E_y$, launching TE 'slab' waveguide modes.

## 5. Spectra of common organic solvents and resonance tuning

In order to determine the best candidates for sensing and strong coupling experiments, we performed FTIR spectroscopy of various solvents for the gratings in this paper. We find that whilst toluene and acetone present very weak absorption bands, both cyclohexane and IPA present relatively strong vibrational bands for analysis in sensing experiments

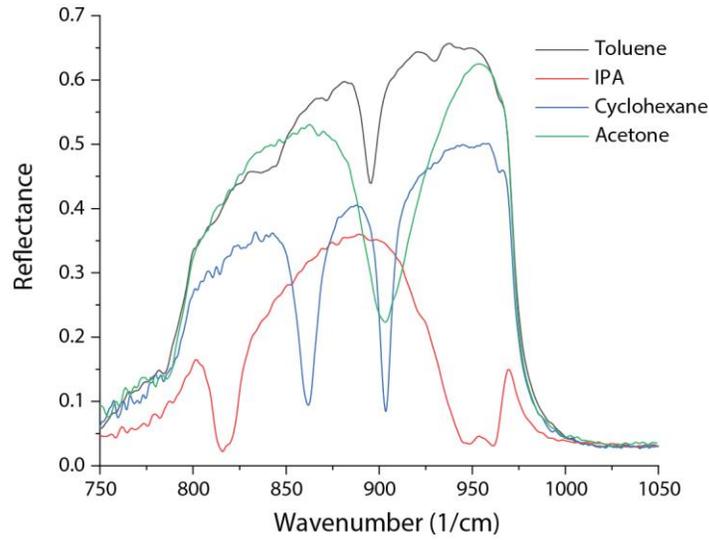

Fig S5. Infrared spectra of solvents taken in the liquid cell next to the gratings (following the process of Fig. 3)

In order to tune the resonance of the $\Lambda=6.49\mu m$ grating in a different solvent to the vibrational band of cyclohexane we use a mixture of IPA and toluene. We find that a 1:1 ratio of toluene and IPA produces an appropriate modal overlap, as shown in Fig. S5

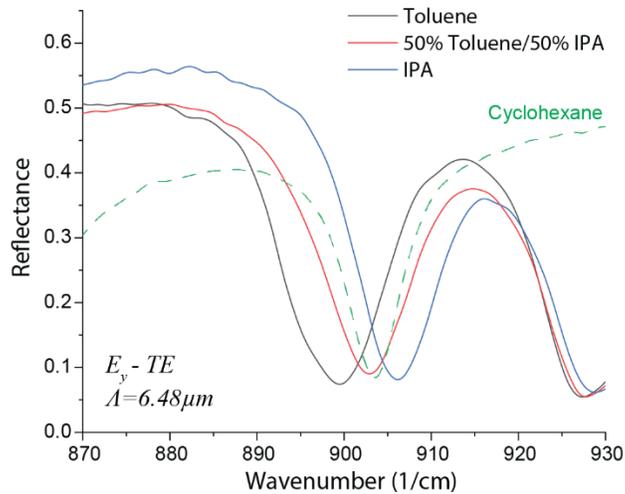

Figure S6. Mode tuning using different solvents for overlapping of cavity mode with the cyclohexane vibrational mode.

## 6. Strong coupling to TE waveguide modes

In order to provide further verification that we are observing a strong coupling effect within our system, we can exploit the analytical model developed in the main text, inputting a complex dielectric function for the liquid as well as for the silicon carbide. To do so we generate a dielectric function for absorption in a polar liquid, using a simple Lorentz oscillator model:

$$\varepsilon(\omega) = \varepsilon_\infty + \frac{S}{\omega_0^2 - \omega^2 - i\gamma\omega}$$

$\varepsilon_\infty$=1.56 is high frequency dielectric constant, $S$=1000cm$^{-2}$ is oscillator strength, $\omega_0$= 903cm$^{-1}$ is vibration energy and $\gamma$=8cm$^{-1}$ is damping. By inputting this dielectric function into the analytical model of the main text we generate the dispersion relation and $Q$ factors shown in Fig. S6. We consider four cases, the grating in air, the grating in a medium with index 1.225, the grating in a medium with a dispersive band at 903cm$^{-1}$ (coincidence with the $l=1$ mode) and a medium with a dispersive band at 939cm$^{-1}$ (coincidence with the $l=3$ mode). From the l=1 case the modes are significantly redshifted when moved into a medium with a higher refractive index. We also see that we can observe a fairly significant mode splitting of 12cm$^{-1}$ for the $l=1$ mode when the vibrational band overlaps with this mode order. This compares favourably to the linewidths extracted from the $Q$ factors, (16cm$^{-1}$ and 9.9cm$^{-1}$) suggesting that it this damping we are on the cusp of a strongly coupled system. Meanwhile, when the vibrational band overlaps with the $l=3$ mode, we observe both a weaker mode splitting of 9cm$^{-1}$ which is unfavourable when compared with the broadened linewidths of 32.2cm$^{-1}$ and 23.35cm$^{-1}$. As a result, this mode would manifest not as a coupled mode, with distinct peaks but as a broadened mode. This highlights the importance of the $l=1$ mode for strong coupling

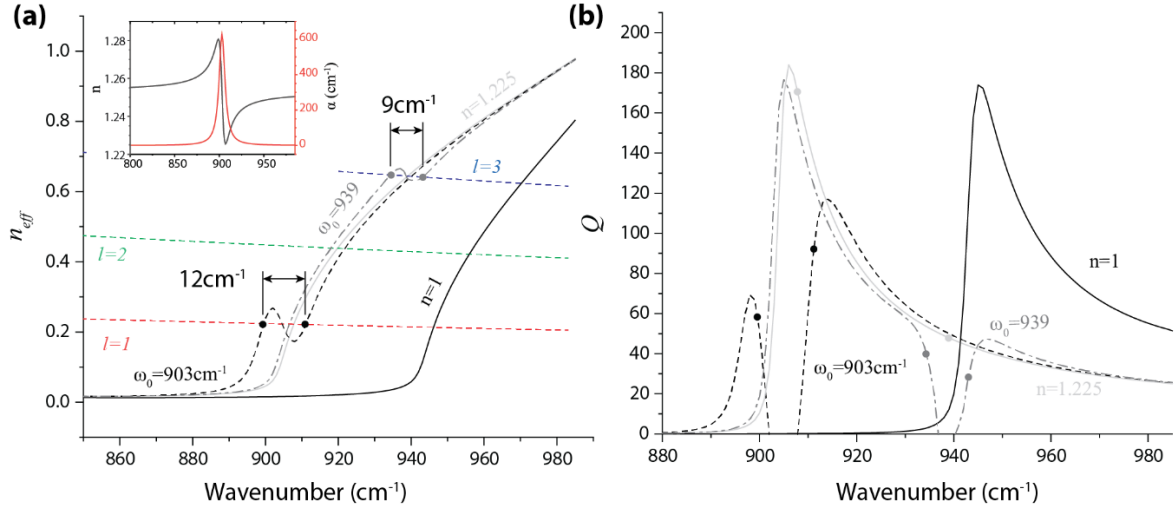

Figure S7. Analytical dispersion relation (a) and $Q$ factors (b) for the TE polarized modes in a $\Lambda$=6.4μm grating. We find that the mode splitting is strongest for the $l=1$ mode, and the $Q$ factors for this mode are significantly higher, making this mode ideal for strong coupling. Inset shows dielectric model for the liquid when $\omega_0$=903cm$^{-1}$.

7. **Study of acetone-IPA mixtures**

In order to assess if these gratings supported a SEIRA effect we studied the influence of small concentrations of IPA on the mode spectrum of the gratings. We consider both deep, and shallow grating modes, and as in the main text, compare the spectrum on the grating, to the control beside the grating. We find that even 2.5% of IPA in acetone produces a measurable dip in the spectrum taken next to the grating. However, on both shallow and deep gratings there is a minimal change in the grating spectrum, suggesting that there is minimal enhancement of absorption from these grating structures in this configuration.

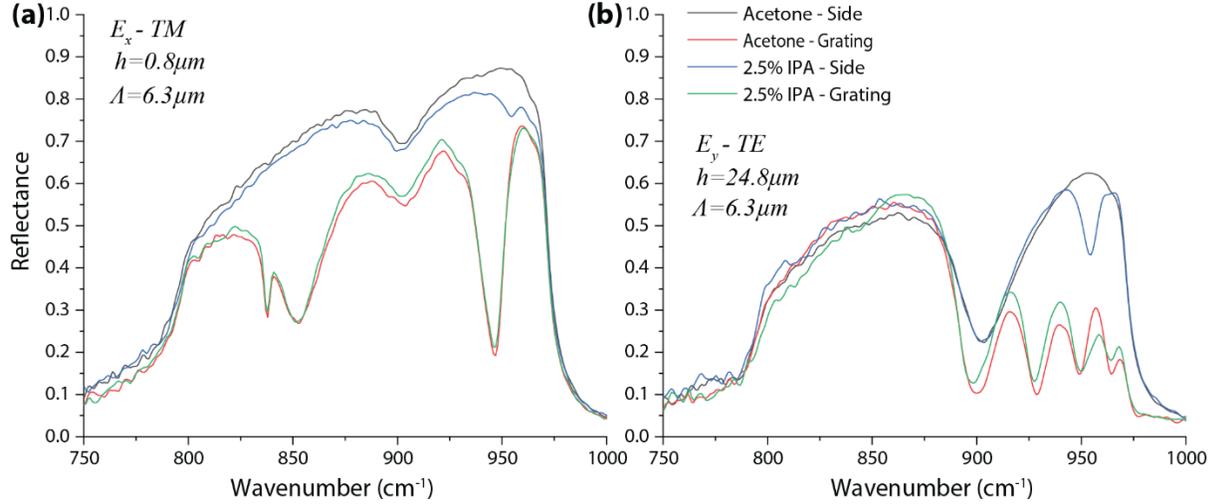

Figure S8. Absence of SEIRA for SiC gratings in acetone/IPA mixtures. Reflectance spectra for (a) shallow and (b) deep gratings in acetone and a 2.5% IPA mixture, compared against spectra next to the grating. For both shallow and deep gratings at 2.5% concentration the absorption line is strong and observable. However, there is minimal changes to the spectra on the gratings.

At higher concentrations of IPA for the shallow grating, we observe a separate IPA absorption peak superimposed on top of the grating mode. Spectra taken on the HAG grating indicate much stronger changes in mode frequency with concentration (Fig. S6 b). In order to extract the tuning range of the polariton mode we use a Lorentz fit to extract the mode positions for the modes indicated in Fig. S6 a/b, shown in Fig. 5c. Modes 1, 2 and 3 redshift with increasing IPA concentration, with a shift of 2.13cm$^{-1}$, 4.3cm$^{-1}$ and 5.1 cm$^{-1}$ respectively. Thus, we can say that the HAG structure shows much stronger sensitivity to the absorption band and the associated change in the dielectric constant of the ambient environment in the acetone IPA mix (see Fig S3).

We also note that at high concentrations of IPA (45%) the mode closest to the IPA line appears to broaden into what could be multiple bands. Whilst in principle this could be strong coupling between the TE waveguide mode, and the IPA absorption line (as observed for cyclohexane), the splitting is not well defined. In order to study if this effect can be attributed to coupling, we can employ our analytical model for the waveguide modes, integrating a dielectric function for our acetone/IPA mixtures (see Fig. S7 for dielectric function). Upon the introduction of the IPA into the dispersion we observe a sigmoidal shape to the dispersion of the mode, which for the correct length cavity will produce a pronounced mode splitting. This contrasts with the much weaker absorption associated with the acetone, which does not produce a mode splitting effect due to its relatively weak and broad absorption. However, the high index of the liquid results in a suppression of the cavity reflectance $R$, resulting overdamping of the mode. This explains the lack of clear strong coupling in acetone-IPA mixtures.

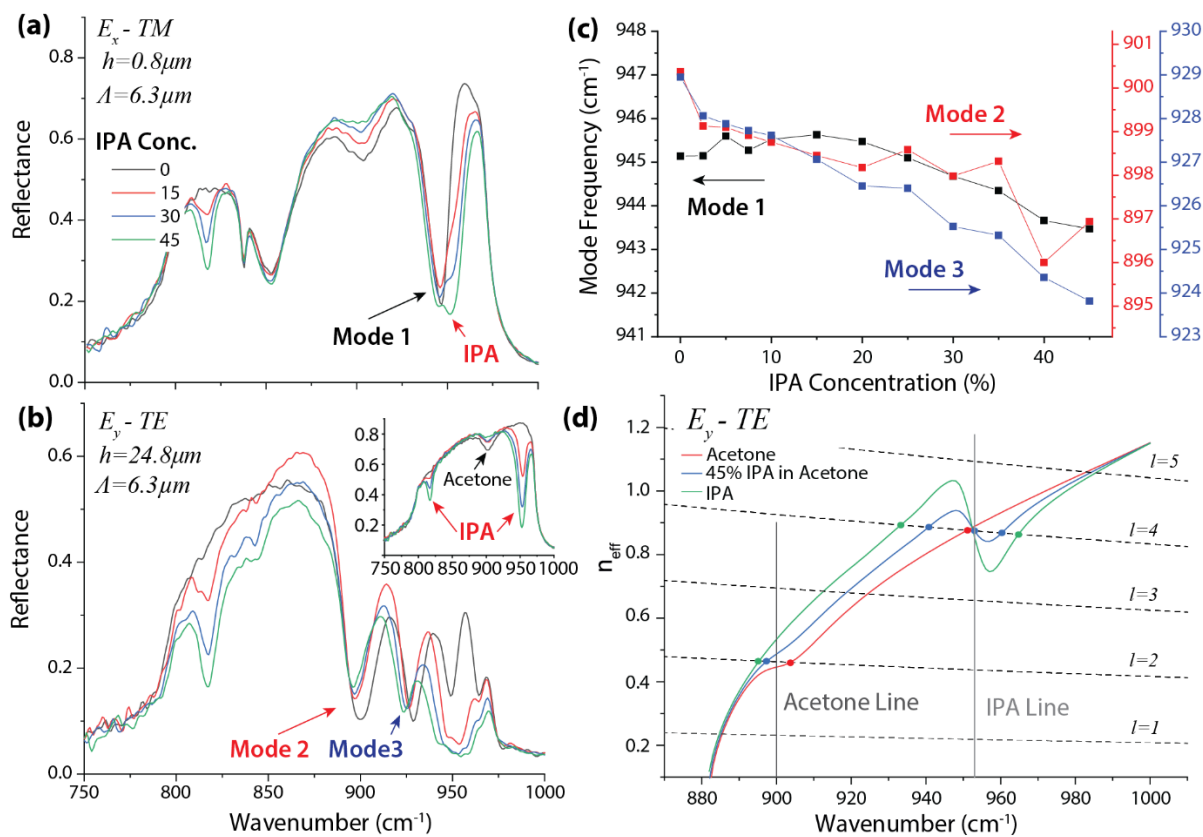

Figure S8. Grating enhanced spectroscopy of acetone-isopropyl alcohol mixtures. (a)/(b) shows the changes in the grating spectra as the concentration of IPA in acetone is increased for a shallow/deep grating respectively. (c) and (d) show the associated frequency tuning of the grating modes as the concentration is increased.

In order to use our analytical model for the waveguide modes to support the hypothesis of strong coupling we use a dielectric function for IPA reported in the literature[17]. We also use the same Lorentz model as above to generate a mode for acetone, where $\varepsilon_\infty=1.88$, $S=2016 cm^{-2}$, $\omega_0 = 900 cm^{-1}$ and $\gamma=20 cm^{-1}$. These values are chosen to be consistent with the amount of absorption observed off the gratings (inset Fig 5b). The dielectric function of mixtures is created by doing a weighted average of the two different materials, with all three shown in Fig S4.

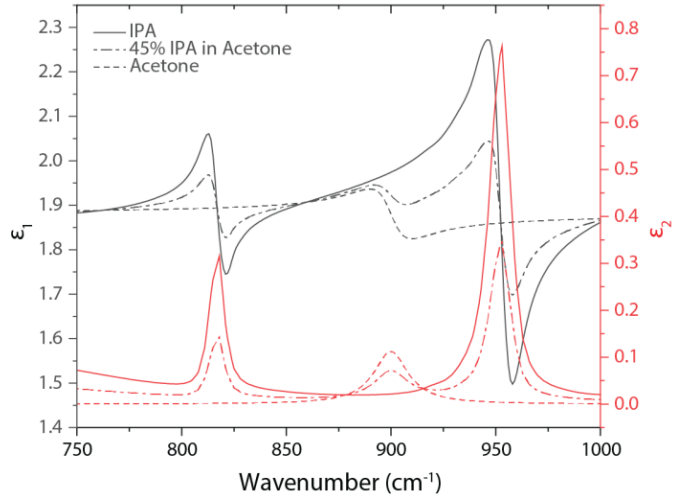

Figure S9. Dielectric function for IPA and acetone in the Reststrahlen band of 4H-SiC used in numerical calculations.

## 8. Determining cavity size in liquid cell

As described in the methods section, our liquid cell experiments produce a small gap between the grating under study and the piece of cover glass. This forms a reflective cavity, producing a series of etalon fringes in optical response, most clearly observed when the sample is in air (Fig. S5). By measuring the spacing of these fringes, we are able to estimate the height of the gap between the cover slip and sample. The frequency spacing of the fringes ($\Delta f$) 673 cm$^{-1}$ (197 cm$^{-1}$), for $h$=0.8µm ($h$=24.8 µm) gratings is related to the size ($L$) of the cavity by $L=c/(2n\Delta f)$, where c is the speed of light and $n$ is the index of the medium. This gives a height of the control cavity of 7.4µm for the grating with $h$=0.8 µm, and 25.4µm for the grating with $h$=24.8 µm.

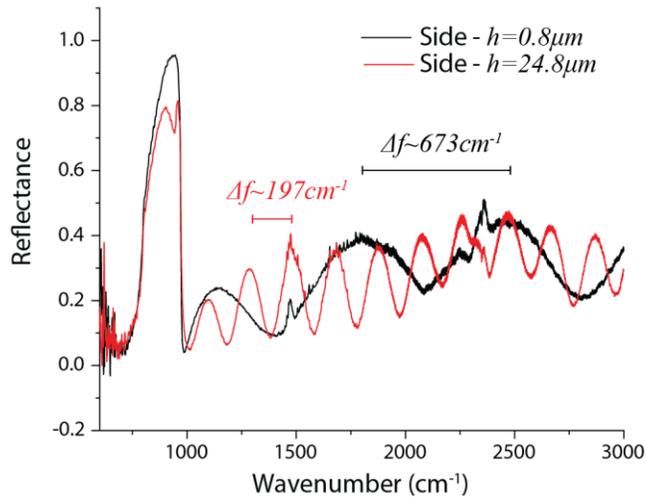

Figure S10. Fringe spacing outside the SiC Reststrahlen band. By analysing the mode spacing on the dry grating we can estimate the gap between the substrate and the BaF$_2$ cover.